\documentclass[fleqn,usenatbib]{mnras}

\usepackage{newtxtext,newtxmath}

\usepackage[T1]{fontenc}

\DeclareRobustCommand{\VAN}[3]{#2}
\let\VANthebibliography\thebibliography
\def\thebibliography{\DeclareRobustCommand{\VAN}[3]{##3}\VANthebibliography}

\usepackage{graphicx}	
\usepackage{amsmath}	

\def\simi   {$\sim$\,}

\def\P     {P$_{1400}$ }

\def\le    {L$_{\rm Edd}$ }

\def \deg         {\text{$^{\circ}$}}

\def \arcsec      {\text{$^{\prime\prime}$}}

\title[RAD12: AGN jet feedback during galaxy merger]{RAD@home citizen science discovery of an AGN spewing a large unipolar radio bubble onto its merging companion galaxy}

\author[Ananda Hota et al.]{
Ananda Hota,$^{1,2}$\thanks{E-mail: hotaananda@gmail.com}
Pratik Dabhade,$^{3,2}$
Sravani Vaddi,$^{4,2}$
Chiranjib Konar,$^{5,2}$
Sabyasachi Pal,$^{6,2}$
Mamta Gulati,$^{7,2}$
\newauthor
C S. Stalin,$^{8,2}$
Avinash Ck,$^{2}$
Avinash Kumar,$^{2}$
Megha Rajoria,$^{2}$
and Arundhati Purohit$^{2}$
\\
$^{1}$\#eAstroLab, UM-DAE Centre for Excellence in Basic Sciences, University of Mumbai, Santacruz-East, Mumbai-400098, India\\
$^{2}$RAD@home Astronomy Collaboratory, Kharghar, Navi Mumbai, PIN 410210, India\\
$^{3}$Observatoire de Paris, LERMA, Coll\`ege de France, PSL University, Sorbonne University, 75014, Paris, France \\
$^{4}$Arecibo Observatory, NAIC, HC3 Box 53995, Arecibo, Puerto Rico, PR 00612, USA \\
$^{5}$Amity Institute of Applied Sciences, Amity University Uttar Pradesh, Sector-125, Noida-201303, India \\
$^{6}$Midnapore City College, Kuturia, Bhadutala, Paschim Medinipur, West Bengal, 721129, India\\
$^{7}$School of Mathematics, Thapar Institute of Engineering and Technology, Patiala, 147004, Punjab, India\\
$^{8}$ Indian Institute of Astrophysics, Koramangala II Block, Bangalore, 560 034, India\\
}

\date{Accepted XXX. Received YYY; in original form ZZZ}

\pubyear{2022}

\begin{document}
\label{firstpage}
\pagerange{\pageref{firstpage}--\pageref{lastpage}}
\maketitle
\begin{abstract}
AGN feedback during galaxy merger has been the most favoured model to explain black hole-galaxy co-evolution. However, how the AGN-driven jet/wind/radiation is coupled with the gas of the merging galaxies, which leads to positive feedback, momentarily enhanced star formation, and subsequently negative feedback, a decline in star formation, is poorly understood. Only a few cases are known where the jet and companion galaxy interaction leads to minor off-axis distortions in the jets and enhanced star formation in the gas-rich minor companions. Here, we briefly report one extraordinary case, RAD12, discovered by RAD@home citizen science collaboratory, where for the first time a radio jet-driven bubble (\simi137 kpc) is showing a symmetric reflection after hitting the incoming galaxy which is not a gas-rich minor but a gas-poor early-type galaxy in a major merger. Surprisingly, neither positive feedback nor any radio lobe on the counter jet side, if any, is detected. It is puzzling if RAD12 is a genuine one-sided jet or a case of radio lobe trapped, compressed and re-accelerated by shocks during the merger. This is the first imaging study of RAD12 presenting follow-up with the GMRT, archival MeerKAT radio data and CFHT optical data. 

\end{abstract}



\section{Introduction}\label{sec:1_intro}
The most popular model of galaxy evolution has been merger and  negative feedback of AGN activity or accretion onto the central supermassive black hole, at the centre of all large galaxies, on the star formation in the surrounding host galaxy \citep{DiMatteo2005,Hopkins2008,McNamara2012,Fabian2012}. Theoretical as well as observational studies  have strongly supported the hypothesis of AGN feedback during the merger of galaxies, leading to relations such as the colour-magnitude diagram, M-$\sigma$ relation, and luminosity function of galaxies \citep{Fabian2012,Heckman14,Terrazas2020}. Majorly there are three mediums of feedback have been identified, namely, intense radiation from the accretion disk, hot/cold material outflow driven outward by the AGN, and radio jets of relativistic magnetised non-thermal plasma \citep{Veilleux2005,Croton06}. Various prescriptions  have been proposed on how to scale the stellar mass of the galaxy to the mass of the black hole, mass of the black hole to the energy output in the form of radiation and then couple it to the interstellar medium (ISM) for a mass of gas lost from the galaxy. The loss or heating of gas in the ISM causes an eventual decrease in star formation of the host galaxy. Several efforts are currently on-going to understand how the radiation/wind/jet is coupled to cold gas in the galaxy so as to predict its effect on quenching of star formation \citep{KH13.M.Sigma,King2015,Dipanjan2016,Cielo2018}. It is essential to investigate the interaction between the radio jet and the ISM of a galaxy. In classical radio galaxies, jet-ISM interaction is hardly observed because the host galaxies are almost always gas-poor ellipticals \citep[see a review by][]{HardcastleCrostron20}. On the other hand, in nearby Seyfert galaxies where the kpc-scale radio jets coincide with the narrow emission line regions (NLR), they are often confusing if the cause is star formation (positive feedback) or AGN-driven shocks and outflows \citep{Nagar1999,Venturi2021}. 

In an ideal and unambiguous case the radio jet hits a neighbouring galaxy and evidence of this jet-galaxy interaction can be found in both radio and optical bands. To date only a few cases are known where the radio jet has been clearly observed to be triggering recent star formation. It is quite likely, that it provides positive feedback (triggers young star formation) on a short time-scale and leads to negative feedback, preventing star formation, which on longer time-scales is required for the understanding of galaxy evolution. We briefly describe five such rare cases of jet-galaxy interactions where enhanced UV and H$\alpha$ emission and distortion of the radio jet have been clearly observed. Understanding the interaction of the radio jet with the stellar and gaseous component of the neighbouring galaxy is important, especially in case of mergers, as it allows to examine the jet-ISM coupling, responsible for positive or negative feedback.

One of the most compelling case of positive feedback is the Minkowski's object \citep{Minkowski1958}, which is a dwarf galaxy located \simi20 kpc north-east of the radio galaxy, NGC541. At the location of interaction the jet develops a radio knot and dwarf galaxy shows evidences of young star formation  \citep{Croft2006,Salome2015}. 3C321 (death star galaxy) is FRII type radio galaxy where its radio jet has been observed to interact with a minor companion galaxy on the western side. It is noted that the jet brightens up in radio and it develops an off-axis diversion to the north-western side extending up to 100 kpc \citep{Evans2008}. 3C34 is a radio galaxy which shows a typical FRII lobe on the west but its eastern lobe is observed to be co-spatial with optically blue and emission line galaxies. It also exhibits double hotspots suggesting jet-galaxy interaction \citep{Best1997}. 3C285 is also a FRII radio galaxy where the eastern jet passes close to a small galaxy with dominant blue colour in optical, which is located  \simi70 kpc away from the radio core \citep{vanBreugel1993,Salome2015}. 3C441 is yet another case where the tip of the north-western FRII lobe appears to be touching a face-on disk/spiral galaxy with the distortion in the radio structure and edge-brightening of the disk galaxy on the side of the lobe suggesting interaction \citep{Lacy1998}. The disk galaxy shows old stellar population consistent with its red colour but also a post-starburst population possibly due to the interaction with the radio lobe. The above-mentioned brief summary of a few rare cases of jet-galaxy interactions clearly demonstrates the effects of positive feedback. If the jet is observed to hit a significantly big galaxy, as in major mergers, we may expect more exotic cases than the ones described above with minor companions. 

In this paper, we present results of RAD12, where we are witnessing a possible case of dry and major merger of two large elliptical galaxies. This brief report on this citizen science discovery is the very first radio-optical imaging study of this object, unveiling a mushroom-shaped radio bubble, unlike any of the above cases.

\begin{figure*}
\centering
\includegraphics[scale=0.18]{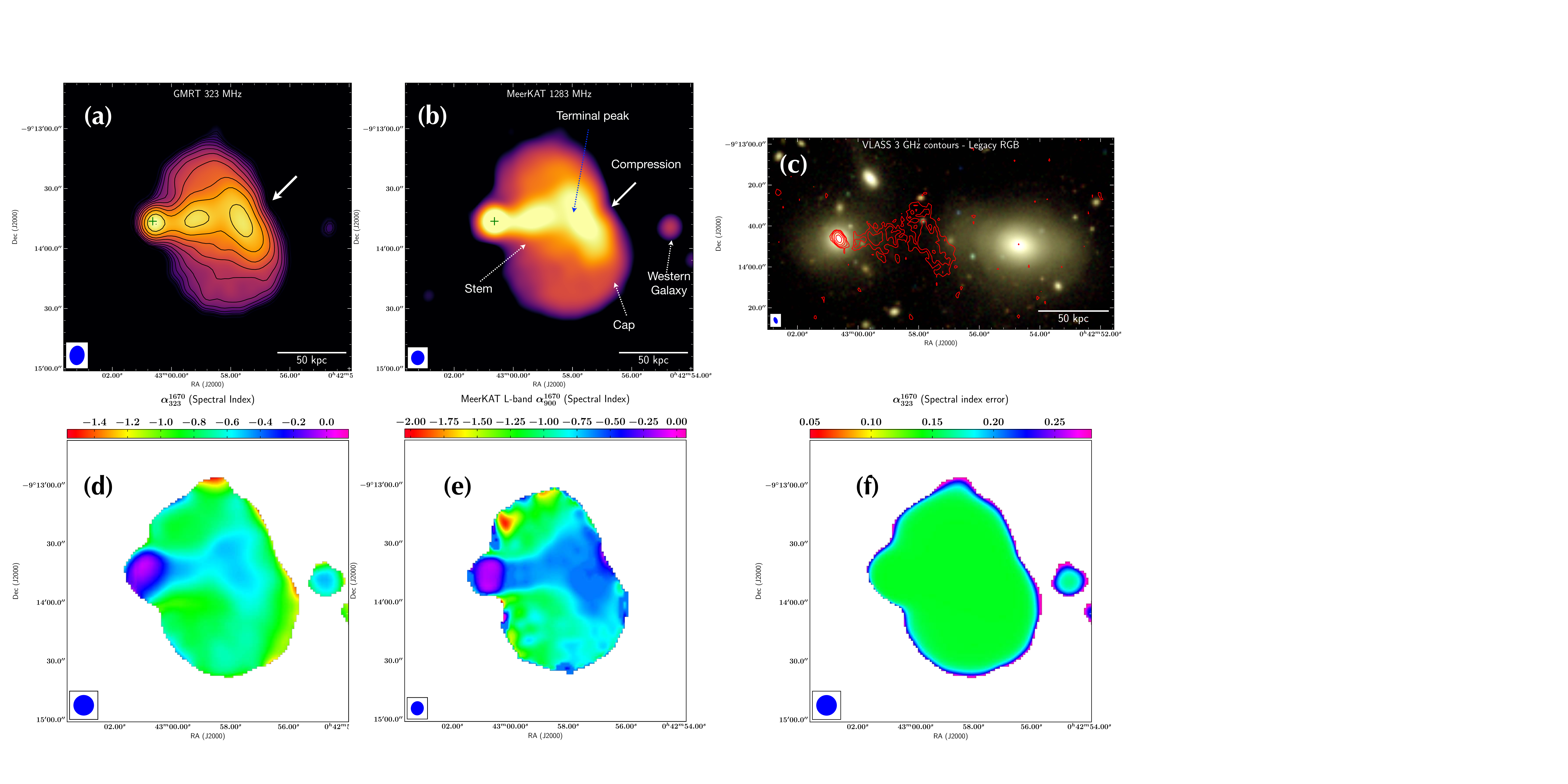}
\caption{RAD12 images: (a): GMRT 323 MHz colour image with contour levels of 0.0007, 0.0008, 0.0009, 0.0011, 0.0014, 0.0018, 0.0025, 0.0036, 0.0053, 0.0080, 0.0124, 0.0193, 0.0302 Jy~beam$^{-1}$, with beam of 10.2\arcsec $\times$ 7.9\arcsec, -6.4\deg. (b): MeerKAT L band image with beam of 7.7\arcsec $\times$ 7.1\arcsec, -8.2\deg. 
(c): Optical colour image from Legacy survey is superposed with 3 GHz contours (levels: 0.0004,0.00066, 0.0013, 0.003, 0.008, 0.022, 0.057, 0.153 Jy~beam$^{-1}$) from VLASS (3.77\arcsec $\times$ 2.29\arcsec, 20.61\deg).
(d): 11\arcsec  $\times$ 11\arcsec, 0\deg spectral index map made using GMRT 323 MHz and MeerKAT L band. (e): In-band spectral index from 900 to 1670 MHz (MeerKAT L band) with beam same as the MeerKAT map. 
 (f): Error in the spectral index map from GMRT and MeerKAT data. Cross mark in green colour in sub-figures a and b marks the position of the optical nucleus of the host galaxy of RAD12. The centre of the western optical galaxy is also detected in both the radio bands.}
\label{fig:ALLIMS}
\end{figure*}

\vspace{-0.7cm}

\section{Observation and Data Analysis}
RAD12 was discovered by RAD@home\footnote{\url{https://www.radathomeindia.org}} citizen science research collaboratory as reported first in \citep{hota14}. The innovative socio-scientific design of the RAD@home collaboratory is briefly described in \citet{Hota16}. It is a zero-funded, zero-infrastructure, nationwide, Inter-University network of professional astronomers and trained citizen scientists (e-/i-astronomers) who are University students or citizens (irrespective of their age and profession) with University-level science education.   

The collaboratory trains eligible citizens to analyse multi-wavelength (UV, Optical, IR, radio) Red Green Blue (RGB)-contour images, and overlays of extragalactic radio sources using simple tools like SAO ds9, NASA- Skyview and RAD@home RGB-maker\footnote{\url{https://www.radathomeindia.org/rgbmaker}}.   

They are trained to find exotic radio sources similar to {\it Speca}, a rare spiral-host episodic radio galaxy \citep{hotaspeca}, and the {\it Cosmic leaf blower galaxy}, NGC3801 which is a sub-galactic scale young radio source in a post-merger early-type galaxy \citep{Hota2012}. Thus, this modified citizen science with multi-wavelength approach has rich and diverse science to offer. Following the above process, a blob of unusual radio emission in the NVSS \citep{nvss} was spotted located between two potential host galaxies, as seen in optical images (e.g DSS), both of which are early-type galaxies. The radio source could not be classified as FRI or FRII type. Higher resolution image from FIRST \citep{beckerfirst95} confirmed the co-incidence of the radio core with the eastern galaxy (SDSS J004300.63--091346.3, \citealt{sdssdr14}) seen in optical and a fainter plume like linear radio emission  was seen only on one side and extending directly towards the western galaxy (SDSS J004254.69--091349.4) which is also at the same redshift ($z$) of 0.076. \citet{Hota16} suggested that it is unlike any head-tail radio galaxies or radio bridges in interacting Taffy galaxies, and the radio jet is likely interacting with the western galaxy but absence of the eastern jet needed further probing. 

Hence, a follow up observation was proposed to the Giant Metrewave Radio Telescope \citep[GMRT;][]{Swarup1991GMRT2} through the proposal titled GOOD-RAC: GMRT Observation of Objects Discovered by RAD@home Astronomy Collaboratory, 32$\_$093 (PI: Ananda Hota). It was observed on 11$^{\rm th}$ September 2017 at 323 MHz with standard procedure of target galaxy observed in between phase and flux density calibrators. The total on source time for the target was 3.33 hours with a bandwidth of 33.33 MHz. The data was analysed using the Source Peeling and Atmospheric Modelling \citep[SPAM;][]{tgss_intema} pipeline.
We also found archival MeerKAT 1283 MHz data for RAD12 as it is present in the field of Abell 85 from the  MeerKAT Galaxy Cluster Legacy Survey \citep[MGCLS;][]{Knowles2022}.
The deep radio images from GMRT and MeerKAT are presented in the top panel of Fig.\ \ref{fig:ALLIMS} (a and b), where the rms of GMRT 323 MHz image is \simi90 $\muup$Jy~beam$^{-1}$ and \simi10 $\muup$Jy~beam$^{-1}$ for MeerKAT L band image. We also created spectral index map of 11\arcsec ~ resolution using GMRT 323 MHz and MeerKAT L band data, which is seen in Fig.\ \ref{fig:ALLIMS}d. The MGCLS data products also provides in-band spectral index map covering the frequency range of 900 to 1670 MHz and it is shown in Fig.\ \ref{fig:ALLIMS}e. 

\vspace{-0.65cm}
\section{Results}
In Fig.\ \ref{fig:ALLIMS},  we present the  GMRT 323 MHz, MeerKAT L band, and their spectral index images for the first time, revealing unprecedented details of RAD12 which are not seen in any other archival radio data. The integrated flux densities of RAD12 at 323 MHz and 1283 MHz are 368.0$\pm$36.8 and 164.4$\pm$8.2 mJy~beam$^{-1}$, respectively. 

The whole radio emission resembles a mushroom, rather a cross section of, and thus various sub-components are named accordingly. 
As seen in Fig.\ \ref{fig:ALLIMS} (panel-a \& b) a plume, poorly collimated jet, like stem of a mushroom, extends from the bright radio core westward. The stem has two minor radio peaks: one at the end, like a hot spot or terminal peak, and the other at the middle, like a knot in the jet. 

The observed stem like feature seen in Fig.\ \ref{fig:ALLIMS} has one-to-one correspondence with the 1.4 GHz image from FIRST as well as 3 GHz VLASS \citep{vlass21} image. The VLASS contours superposed on optical colour image from Legacy survey\footnote{\url{https://www.legacysurvey.org}} database demonstrate absence of any compact jet or hotspot but a diffuse emission structure of the whole stem. However, the feature resembling a cap of the mushroom is seen, both to the north and south of the stem, in GMRT and MeerKAT images, but not detected in FIRST and VLASS (see Fig.\ \ref{fig:ALLIMS}b for name of sub-structures.
The stem shows a compression at the western edge, after the terminal peak, and the emission diffuses towards the north and south forming the cap (total N-S extent 137 kpc or 95\arcsec).  The GMRT and MeerKAT contours are closely spaced on the western side of the terminal peak than on its north and south sides, suggesting an abrupt stoppage of plasma, especially in the absence of any compact feature in the VLASS image at this location.
This compression is a subtle feature and may require further higher resolution and polarisation observations for further probing. The diffuse emission of the cap, resembles back flow observed often seen in typical FRII type radio lobes, extends to the east reaching almost close to the radio core. However, the features in this radio mushroom have subtle differences from FRII lobes, which is discussed later in the paper.

The spectral index maps (Fig.\ \ref{fig:ALLIMS} d \& e) display smooth variations along the expected dynamical features, where we observe the core to be very flat, the stem $\alpha$ $\sim$ --0.5, and the region beyond terminal peak or compressed region has $\alpha$ more flatter than the overall stem. The diffuse emission in the cap, as expected, is steeper with $\alpha$ ranging from $\sim$ --0.8 to $\sim$ --1.5. Note that the errors on the spectral index, presented in the same Figure, is around 0.15, which is much smaller than differences found in different sub-structures. This gradual variation of spectral index from core to stem, to terminal peak, and diffuse region of the cap show perfect textbook example of fresh plasma supply, re-acceleration at the hot-spot and steepening in the back flow of a typical FRII radio galaxy lobe. However, there are two major differences, 1) instead of narrow particle beam or jet and compact hotspot structures, RAD12 stem is not very well collimated (at least on kpc scales) and 2. the terminal peak of the stem or the compressed region, although shows spectral flattening, shows a concave structure (indicated with an arrow in Fig.\ \ref{fig:ALLIMS}) instead of a convex structure or a bowshock structure, which is typically seen in FRII radio galaxy lobes.

In Fig.\ \ref{fig:cfht} the GMRT 323 MHz radio contours are overlaid on a deep (exposure time = 5140 secs) r-band optical image obtained from the CFHT archive (MegaPrime instrument on 3.6m telescope). While the radio core coincide with the nucleus of the eastern galaxy (SDSS J004300.63--091346.3) hereafter host galaxy, a faint point radio source can be identified to coincide with the larger and brighter western galaxy (SDSS J004254.69--091349.4) hereafter companion galaxy. The spectral index of this point source from GMRT and MeerKAT data is $\alpha$ $\sim$ --0.5 (see Fig.\ \ref{fig:ALLIMS}c) confirming its association with the nucleus of the companion. The spectroscopic redshift of the host and companion galaxies are 0.07628$\pm$0.00001 and 0.07618$\pm$0.00001, respectively. 
To the north of the eastern galaxy, a relatively smaller galaxy (SDSS J004259.59--091316.8) with a faint but curved stellar tail extending northward is seen. The northern galaxy has a photometric redshift of 0.081$\pm$0.009 and is quite likely a second merging companion of RAD12. Although there are many faint foreground/background galaxies in the region around RAD12 and its western companion, the middle and terminal radio peaks on the stem does not show any optical counterpart. The radio-optical overlay as seen in Fig.\ \ref{fig:cfht}, the stem seems to terminate, with steep gradient of radio contours, in a region touching (in projection) the outer boundary of the companion galaxy. Furthermore, the terminal peak has north-south orientation unlike the middle peak on the stem. Thus, the radio mushroom seem to inflate to both north and south direction relatively easily after getting obstructed possibly by the gaseous halo or stellar shells (Fig.\ \ref{fig:ALLIMS}c and Fig.\ \ref{fig:cfht}) of the companion galaxy or the gas in the orbital plane of the galaxy merger. 

\vspace{-0.946cm}
\section{Discussion}
Below we present discussion on a few possible scenarios explaining the observed radio-optical features of RAD12 and its companion galaxy.

{\bf Jet-galaxy interaction in a Major merger:} 
RAD12 is unlike any of the five jet-galaxy interactions described in the introduction where radio jet hit only a minor gas-rich merging companion. This is possibly the first clear case of a major merger, where the companion galaxy, which is being hit by the jet, is a gas-poor bigger and brighter elliptical galaxy. In other examples of jet-galaxy interactions, minor distortions (brightening of a knot, off axis diversions, double hotspot) were observed but RAD12 is the first case of complete halt of the jet with structural depression and spectral flattening. Most significantly it is a symmetric and possibly experiencing complete reflection forming a mushroom-shaped bubble. The stem is 64 kpc long (core to terminal peak is 45\arcsec ) which is nearly half of the distance between nuclei, 128 kpc (88\arcsec). It is puzzling what is in the middle of this gas-poor merging ellipticals that push the jet backward. Is it the bowshock of a major merger or magnetosphere of the incoming companion galaxy or more exotic things like dark matter?

\begin{figure}
    \centering
    \includegraphics[scale=0.132]{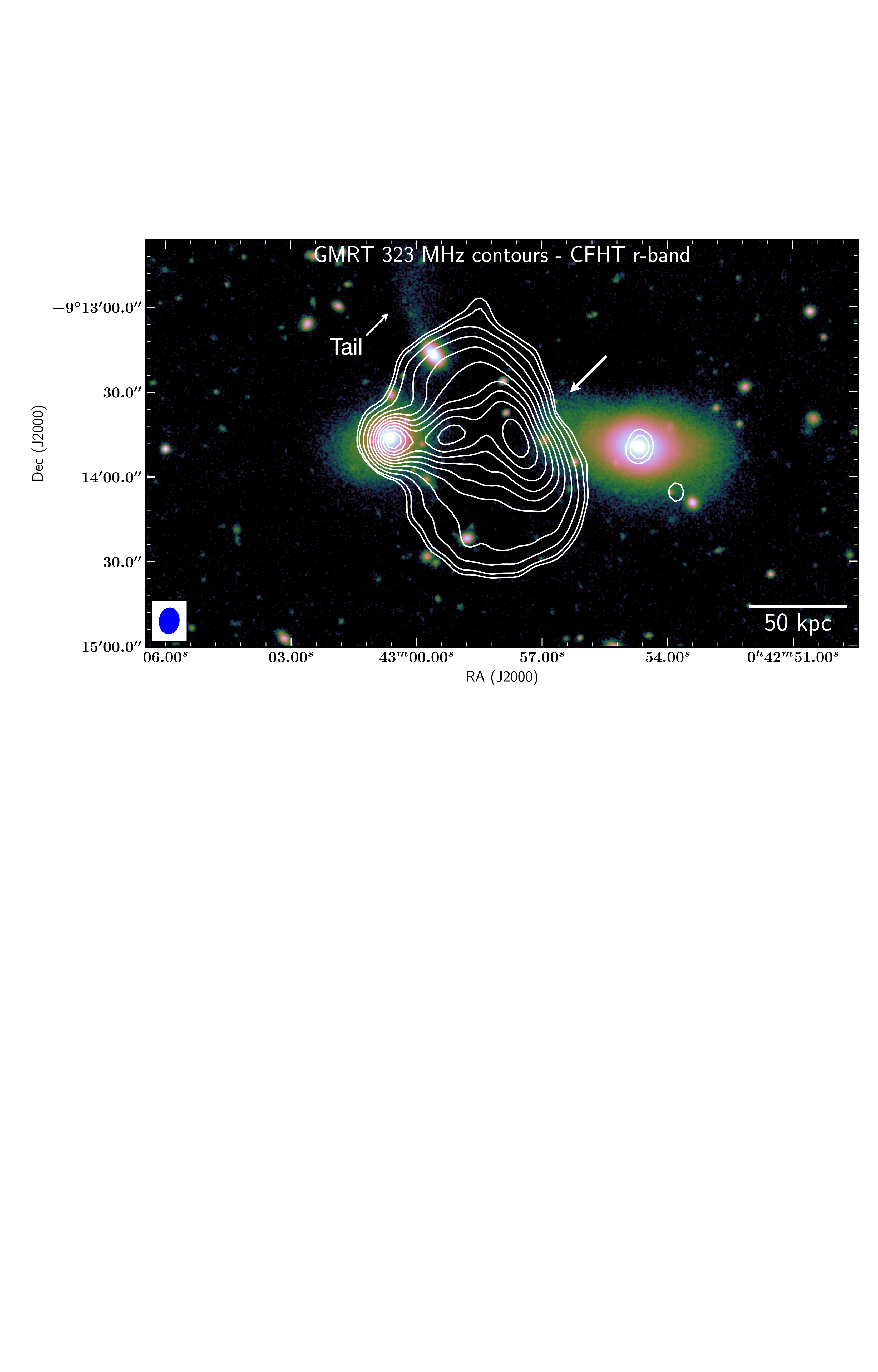}
    \caption{Optical r-band image from the CFHT of galaxies SDSS J004300.63--091346.3, host of RAD12 (left) and companion galaxy  SDSS J004254.69--091349.4 (right) overlaid with GMRT 323 MHz contours.}
    \label{fig:cfht}
\end{figure}

{\bf No sign of positive feedback:}
Unlike the five known cases where the jet-galaxy interaction triggers young star formation showing up in UV-optical colours, RAD12 does not seem to show any. Nothing is seen at the terminal peak or contour suppression either in individual bands of SDSS or in colour image. The u and r band SDSS magnitudes of the host galaxy are u (18.21 $\pm$ 0.03) and r (15.36 $\pm$ 0.00), and that for the brighter ($\sim$ 1 mag.) companion galaxy are u (17.23 $\pm$ 0.03), r (14.22 $\pm$ 0.00). Thus, the u-r colours, 2.85 and 3.01, put both of them in the typical red and dead elliptical galaxy category. It is possible that the incoming galaxy is too big or gas-poor for the jet to trigger visible galaxy-wide young star formation. Interestingly, the deep CFHT optical image shows a large region (\simi25 $\times$ 25 kpc) of faint diffuse optical emission observed close to the region of compressed radio contours (indicated with an arrow in Fig.\ \ref{fig:cfht}). Such optical emission is not observed on the diametrically opposite side of the companion galaxy. Causal connection between compressed radio emission and this asymmetric diffuse emission in optical is unclear at the moment. Note that a small galaxy (SDSS J004256.93--091347.1) seen in the region of compression is a background galaxy ($z_{\rm photo}$= 0.220$\pm$0.108). 

{\bf Unlike a Head-Tail radio galaxy:} The core-stem structure resembles a head-tail radio galaxy, typically seen in galaxy clusters. The observed bright and unresolved core component could possibly harbour a narrow-angle tailed radio source. However, based on our data we observe that the core is extremely flat, stem is also flat with freshly supplied plasma from the core, and shows no sign of bifurcation  (e.g. VLASS image) or spectral steepening in the stem. Hence, our observational evidence disfavours the possibility of a narrow-angle tailed radio source. Furthermore, the creation of a big radio bubble with its spectrally distinct features encompassing the stem by a head-tail radio galaxy is difficult to explain.

{\bf Intrinsically One-sided AGN radio jet:} The L-band luminosity of the radio source is  2.3 $\times$ 10$^{24}$ W Hz$^{-1}$. To put this radio loud AGN in to perspective, one can compare its location in the power vs. projected size plot (P-D diagram) of all classes of radio sources (see Fig. 2 of \citealt{HardcastleCrostron20}). RAD12 is  located in between mean location of radio quiet/weak quasars and FRI radio galaxies suggesting almost similar power but smaller jet size. Hence, detection of only one possible radio lobe is puzzling. In the MeerKAT image, the flux density of the whole mushroom structure is \simi140 mJy (total 164 mJy - core flux density of 24mJy) and the 3$\sigma$ upper limit of the non-detection for the western counterpart is \simi0.03 mJy. This gives an asymmetry by a factor of about 4660 which is extremely unusual for asymmetric radio galaxies \citep{GK04}. Brightness asymmetry in jets can arise due to Doppler boosting, so independent clues on AGN orientation was looked for. Faint and narrow emission lines seen in SDSS spectrum and its location in the IR colour-colour plot \citep[e.g.][]{Gurkan2014} suggest clearly that the target is a low excitation radio galaxy and not a quasar. In other words, the radio-jet in RAD12 does not seem to be oriented close to the line of sight. Hence, this apparently missing eastern emission in any radio data presents a strong case for intrinsically one-sided launching of large-scale (50 kpc) radio jet. Doppler boosting can be attributed to only VLBI-scale (pc) jet-counter jet asymmetry and environmental asymmetry in such early-type galaxy is not expected before the jet hits the ISM of the companion galaxy $\sim$ 60 kpc away. Hence, the stem appears to be intrinsically asymmetric with poorly collimated jet feature which could have grown to a bubble or mushroom structure only after interaction with the companion galaxy. Further detailed observations are required to confirm if RAD12 indeed has an one-sided jet.       

{\bf Merger Trapped radio bubble:} Alternative to the above one-sided jet model, we propose that RAD12 is representing a rare/transient `caught in the act' phase of merging galaxies where one of the radio lobe is trapped in between two merging nuclei. In such a case, the  radio lobe on the counter jet side (eastern) escapes outward to a low-density environment and can miss detection due to efficient diffusion. Due to dense inter-nuclei medium and/or shocks in between approaching nuclei, the brightness of the trapped lobe can be enhanced or diffuse less-efficiently. Adiabatic compression and/shock re-acceleration of this radio plasma can be compared to that observed in a radio ``Phoenix'' in clusters of galaxies \citep{EnsslinGK01}. 

Such well-shaped radio bubbles are known in low-power radio sources hosted in both spiral and elliptical galaxies. The radio luminosity of RAD12 is typical of a representative sample of radio bubbles in the nearby Universe compiled by \citet{hotasaikia06}. Beautiful example of such a radio bubble is seen in Seyfert-Starburst composite NGC6764, where the poorly collimated nuclear jet interacts with the gas disk of the spiral galaxy and then expands laterally to form bipolar radio bubbles along the rotation axis of the disk \citep{hotasaikia06}. Such bubbles can drive atomic, molecular and ionised gas outflows. Follow up observations with {\it Chandra} showed the radio bubble to have perfect correspondence in X-ray suggesting supersonic expansion of the bubble driven shock waves and heating of the surrounding ISM. The total energy of the X-ray bubble, 10$^{56}$ ergs, is not only comparable to that in low-power radio galaxies but also more than kinetic energy of the outflowing cold gas \citep[e.g.][]{Croston08}. Interestingly, the dynamical age of this radio bubble also coincide with a break in the nuclear star formation history \citep{Hota2012}. Future study of many such intra-nuclei radio bubbles can reveal thermostat like nature of AGN feedback-fueling cycles or when the bubble is powerful enough to drive outflow and when it is weak to suggest trapped bubble or inflow of gas to feed/fuel the AGN/nuclear star formation. 

{\bf Future Direction:} 
Using the available radio and optical archival data, it is possible to find objects such as RAD12 and which upon further follow up observations can be studied in depth to unravel the underlying physics involved in the AGN feedback coupled with galaxy mergers. The GMRT and MeerKAT data clearly demonstrate the uniqueness of RAD12 and justifies further examination using multi-wavelength data, e.g. radio polarisation, X-ray imaging and spectroscopy of ionised/molecular/atomic gas observations. Particularly, studying the gas kinematics can possibly shed more light on the interaction of a radio bubble with the ISM of a merging galaxy system and thereby improving our understanding of AGN feedback.

Detection of large radio jets are no longer confined to classical ellipticals but also found in many spirals and narrow-line Seyfert 1 galaxies which are often disk galaxies \citep[e.g][]{hotaspeca,Rakshit2018,Webster2021}.
Hence, our current understanding of the prevalence of jet-mode feedback possibly requires significant upgrade. 
Similarly, as this unusual radio source could not have been correctly characterised from the NVSS, TGSS \citep{tgss_intema}, DSS like all sky survey data alone, it is likely that such radio-mode feedbacks with non-standard radio jets and unusual spectral nature may be more prevalent in galaxies which does not even look like merging/interacting. Ongoing surveys like VLASS \citep{vlass21} at 3 GHz and LoTSS at 144 MHz \citep{lotssdr2} are ideal for citizen science projects to discover more such cases of AGN jet feedback in merging galaxies. Since the age of relics of past AGN activities in radio bands could be 100 million years or more compared to about 10,000 years in optical band \citep[e.g. Hanny's Voorworp;][]{Lintott09,Keel2012}, 
new sensitive radio surveys will unearth significant samples. Eventually in the era of Square Kilometre Array radio telescope, ``smoking gun'' evidences of AGN feedback in a sample can be found where age of the radio relic could possibly match the time-scale of decline in star formation in a post-merger early-type galaxy \citep{Hota16}. 

\vspace{-0.7cm}
\section*{Acknowledgements}

We thank the anonymous referee for constructive comments on the manuscript. We thank the staff of the GMRT that made these observations possible. GMRT is run by the National Centre for Radio Astrophysics of the Tata Institute of Fundamental Research. 
Based on observations obtained with MegaPrime/MegaCam, a joint project of CFHT and CEA/DAPNIA, at the Canada-France-Hawaii Telescope (CFHT) which is operated by the National Research Council (NRC) of Canada, the Institut National des Science de l'Univers of the Centre National de la Recherche Scientifique (CNRS) of France, and the University of Hawaii. The observations at the Canada-France-Hawaii Telescope were performed with care and respect from the summit of Maunakea which is a significant cultural and historic site. 
AH acknowledges University Grants Commission (UGC, Ministry of Education, Govt. of India) for monthly salary. 

The long list of national and international organisations which have helped the establishment and growth of the first Indian astronomical citizen science research platform RAD@home are acknowledged in detail at \url{https://radathomeindia.org/brochure}. 

\vspace{-0.7cm}

 \section*{Data Availability}
 \begin{small}
The GMRT data presented in this article can be accessed from the GMRT Online Archive\footnote{ https://naps.ncra.tifr.res.in/goa/data/search}.
The derived data generated in this research will be shared on reasonable request to the corresponding author.
\end{small}
\vspace{-0.7cm}
\bibliographystyle{mnras}
\bibliography{ref} 

\begin{thebibliography}{}
\makeatletter
\relax
\def\mn@urlcharsother{\let\do\@makeother \do\$\do\&\do\#\do\^\do\_\do\%\do\~}
\def\mn@doi{\begingroup\mn@urlcharsother \@ifnextchar [ {\mn@doi@}
  {\mn@doi@[]}}
\def\mn@doi@[#1]#2{\def\@tempa{#1}\ifx\@tempa\@empty \href
  {http://dx.doi.org/#2} {doi:#2}\else \href {http://dx.doi.org/#2} {#1}\fi
  \endgroup}
\def\mn@eprint#1#2{\mn@eprint@#1:#2::\@nil}
\def\mn@eprint@arXiv#1{\href {http://arxiv.org/abs/#1} {{\tt arXiv:#1}}}
\def\mn@eprint@dblp#1{\href {http://dblp.uni-trier.de/rec/bibtex/#1.xml}
  {dblp:#1}}
\def\mn@eprint@#1:#2:#3:#4\@nil{\def\@tempa {#1}\def\@tempb {#2}\def\@tempc
  {#3}\ifx \@tempc \@empty \let \@tempc \@tempb \let \@tempb \@tempa \fi \ifx
  \@tempb \@empty \def\@tempb {arXiv}\fi \@ifundefined
  {mn@eprint@\@tempb}{\@tempb:\@tempc}{\expandafter \expandafter \csname
  mn@eprint@\@tempb\endcsname \expandafter{\@tempc}}}

\bibitem[\protect\citeauthoryear{{Abolfathi} et~al.,}{{Abolfathi}
  et~al.}{2018}]{sdssdr14}
{Abolfathi} B.,  et~al., 2018, \mn@doi [ApJS] {10.3847/1538-4365/aa9e8a}, \href
  {https://ui.adsabs.harvard.edu/\#abs/2018ApJS..235...42A} {235, 42}

\bibitem[\protect\citeauthoryear{{Becker}, {White}  \& {Helfand}}{{Becker}
  et~al.}{1995}]{beckerfirst95}
{Becker} R.~H.,  {White} R.~L.,   {Helfand} D.~J.,  1995, \mn@doi [\apj]
  {10.1086/176166}, \href {http://adsabs.harvard.edu/abs/1995ApJ...450..559B}
  {450, 559}

\bibitem[\protect\citeauthoryear{{Best}, {Longair}  \& {Roettgering}}{{Best}
  et~al.}{1997}]{Best1997}
{Best} P.~N.,  {Longair} M.~S.,   {Roettgering} H.~J.~A.,  1997, \mn@doi
  [\mnras] {10.1093/mnras/292.4.758}, \href
  {https://ui.adsabs.harvard.edu/abs/1997MNRAS.292..758B} {292, 758}

\bibitem[\protect\citeauthoryear{{Cielo}, {Bieri}, {Volonteri}, {Wagner}  \&
  {Dubois}}{{Cielo} et~al.}{2018}]{Cielo2018}
{Cielo} S.,  {Bieri} R.,  {Volonteri} M.,  {Wagner} A.~Y.,   {Dubois} Y.,
  2018, \mn@doi [\mnras] {10.1093/mnras/sty708}, \href
  {https://ui.adsabs.harvard.edu/abs/2018MNRAS.477.1336C} {477, 1336}

\bibitem[\protect\citeauthoryear{{Condon}, {Cotton}, {Greisen}, {Yin},
  {Perley}, {Taylor}  \& {Broderick}}{{Condon} et~al.}{1998}]{nvss}
{Condon} J.~J.,  {Cotton} W.~D.,  {Greisen} E.~W.,  {Yin} Q.~F.,  {Perley}
  R.~A.,  {Taylor} G.~B.,   {Broderick} J.~J.,  1998, \mn@doi [\aj]
  {10.1086/300337}, \href {http://adsabs.harvard.edu/abs/1998AJ....115.1693C}
  {115, 1693}

\bibitem[\protect\citeauthoryear{{Croft} et~al.,}{{Croft}
  et~al.}{2006}]{Croft2006}
{Croft} S.,  et~al., 2006, \mn@doi [\apj] {10.1086/505526}, \href
  {https://ui.adsabs.harvard.edu/abs/2006ApJ...647.1040C} {647, 1040}

\bibitem[\protect\citeauthoryear{{Croston}, {Hardcastle}, {Kharb}, {Kraft}  \&
  {Hota}}{{Croston} et~al.}{2008}]{Croston08}
{Croston} J.~H.,  {Hardcastle} M.~J.,  {Kharb} P.,  {Kraft} R.~P.,   {Hota} A.,
   2008, \mn@doi [\apj] {10.1086/592268}, \href
  {https://ui.adsabs.harvard.edu/abs/2008ApJ...688..190C} {688, 190}

\bibitem[\protect\citeauthoryear{{Croton} et~al.,}{{Croton}
  et~al.}{2006}]{Croton06}
{Croton} D.~J.,  et~al., 2006, \mn@doi [\mnras]
  {10.1111/j.1365-2966.2005.09675.x}, \href
  {https://ui.adsabs.harvard.edu/abs/2006MNRAS.365...11C} {365, 11}

\bibitem[\protect\citeauthoryear{{Di Matteo}, {Springel}  \& {Hernquist}}{{Di
  Matteo} et~al.}{2005}]{DiMatteo2005}
{Di Matteo} T.,  {Springel} V.,   {Hernquist} L.,  2005, \mn@doi [\nat]
  {10.1038/nature03335}, \href
  {https://ui.adsabs.harvard.edu/abs/2005Natur.433..604D} {433, 604}

\bibitem[\protect\citeauthoryear{{En{\ss}lin} \& {Gopal-Krishna}}{{En{\ss}lin}
  \& {Gopal-Krishna}}{2001}]{EnsslinGK01}
{En{\ss}lin} T.~A.,  {Gopal-Krishna} 2001, \mn@doi [\aap]
  {10.1051/0004-6361:20000198}, \href
  {https://ui.adsabs.harvard.edu/abs/2001A&A...366...26E} {366, 26}

\bibitem[\protect\citeauthoryear{{Evans} et~al.,}{{Evans}
  et~al.}{2008}]{Evans2008}
{Evans} D.~A.,  et~al., 2008, \mn@doi [\apj] {10.1086/527410}, \href
  {https://ui.adsabs.harvard.edu/abs/2008ApJ...675.1057E} {675, 1057}

\bibitem[\protect\citeauthoryear{{Fabian}}{{Fabian}}{2012}]{Fabian2012}
{Fabian} A.~C.,  2012, \mn@doi [\araa] {10.1146/annurev-astro-081811-125521},
  \href {https://ui.adsabs.harvard.edu/abs/2012ARA&A..50..455F} {50, 455}

\bibitem[\protect\citeauthoryear{{Gopal-Krishna} \& {Wiita}}{{Gopal-Krishna} \&
  {Wiita}}{2004}]{GK04}
{Gopal-Krishna} {Wiita} P.~J.,  2004, arXiv e-prints, \href
  {https://ui.adsabs.harvard.edu/abs/2004astro.ph..9761G} {pp
  astro--ph/0409761}

\bibitem[\protect\citeauthoryear{{Gordon} et~al.,}{{Gordon}
  et~al.}{2021}]{vlass21}
{Gordon} Y.~A.,  et~al., 2021, \mn@doi [\apjs] {10.3847/1538-4365/ac05c0},
  \href {https://ui.adsabs.harvard.edu/abs/2021ApJS..255...30G} {255, 30}

\bibitem[\protect\citeauthoryear{{G{\"u}rkan}, {Hardcastle}  \&
  {Jarvis}}{{G{\"u}rkan} et~al.}{2014}]{Gurkan2014}
{G{\"u}rkan} G.,  {Hardcastle} M.~J.,   {Jarvis} M.~J.,  2014, \mn@doi [\mnras]
  {10.1093/mnras/stt2264}, \href
  {https://ui.adsabs.harvard.edu/abs/2014MNRAS.438.1149G} {438, 1149}

\bibitem[\protect\citeauthoryear{{Hardcastle} \& {Croston}}{{Hardcastle} \&
  {Croston}}{2020}]{HardcastleCrostron20}
{Hardcastle} M.~J.,  {Croston} J.~H.,  2020, \mn@doi [\nar]
  {10.1016/j.newar.2020.101539}, \href
  {https://ui.adsabs.harvard.edu/abs/2020NewAR..8801539H} {88, 101539}

\bibitem[\protect\citeauthoryear{{Heckman} \& {Best}}{{Heckman} \&
  {Best}}{2014}]{Heckman14}
{Heckman} T.~M.,  {Best} P.~N.,  2014, \mn@doi [\araa]
  {10.1146/annurev-astro-081913-035722}, \href
  {https://ui.adsabs.harvard.edu/abs/2014ARA&A..52..589H} {52, 589}

\bibitem[\protect\citeauthoryear{{Hopkins}, {Hernquist}, {Cox}  \&
  {Kere{\v{s}}}}{{Hopkins} et~al.}{2008}]{Hopkins2008}
{Hopkins} P.~F.,  {Hernquist} L.,  {Cox} T.~J.,   {Kere{\v{s}}} D.,  2008,
  \mn@doi [\apjs] {10.1086/524362}, \href
  {https://ui.adsabs.harvard.edu/abs/2008ApJS..175..356H} {175, 356}

\bibitem[\protect\citeauthoryear{{Hota} \& {Saikia}}{{Hota} \&
  {Saikia}}{2006}]{hotasaikia06}
{Hota} A.,  {Saikia} D.~J.,  2006, \mn@doi [\mnras]
  {10.1111/j.1365-2966.2006.10738.x}, \href
  {https://ui.adsabs.harvard.edu/abs/2006MNRAS.371..945H} {371, 945}

\bibitem[\protect\citeauthoryear{{Hota} et~al.,}{{Hota}
  et~al.}{2011}]{hotaspeca}
{Hota} A.,  et~al., 2011, \mn@doi [\mnras] {10.1111/j.1745-3933.2011.01115.x},
  \href {http://adsabs.harvard.edu/abs/2011MNRAS.417L..36H} {417, L36}

\bibitem[\protect\citeauthoryear{{Hota}, {Rey}, {Kang}, {Kim}, {Matsushita}  \&
  {Chung}}{{Hota} et~al.}{2012}]{Hota2012}
{Hota} A.,  {Rey} S.-C.,  {Kang} Y.,  {Kim} S.,  {Matsushita} S.,   {Chung} J.,
   2012, \mn@doi [\mnras] {10.1111/j.1745-3933.2012.01231.x}, \href
  {https://ui.adsabs.harvard.edu/abs/2012MNRAS.422L..38H} {422, L38}

\bibitem[\protect\citeauthoryear{{Hota} et~al.,}{{Hota} et~al.}{2014}]{hota14}
{Hota} A.,  et~al., 2014, in Astronomical Society of India Conference Series.
  pp 141--145 (\mn@eprint {arXiv} {1402.3674})

\bibitem[\protect\citeauthoryear{{Hota} et~al.,}{{Hota} et~al.}{2016}]{Hota16}
{Hota} A.,  et~al., 2016, \mn@doi [Journal of Astrophysics and Astronomy]
  {10.1007/s12036-016-9415-8}, \href
  {https://ui.adsabs.harvard.edu/abs/2016JApA...37...41H} {37, 41}

\bibitem[\protect\citeauthoryear{{Intema}, {Jagannathan}, {Mooley}  \&
  {Frail}}{{Intema} et~al.}{2017}]{tgss_intema}
{Intema} H.~T.,  {Jagannathan} P.,  {Mooley} K.~P.,   {Frail} D.~A.,  2017,
  \mn@doi [\aap] {10.1051/0004-6361/201628536}, \href
  {http://adsabs.harvard.edu/abs/2017A%26A...598A..78I} {598, A78}

\bibitem[\protect\citeauthoryear{{Keel} et~al.,}{{Keel}
  et~al.}{2012}]{Keel2012}
{Keel} W.~C.,  et~al., 2012, \mn@doi [\mnras]
  {10.1111/j.1365-2966.2011.20101.x}, \href
  {https://ui.adsabs.harvard.edu/abs/2012MNRAS.420..878K} {420, 878}

\bibitem[\protect\citeauthoryear{{King} \& {Pounds}}{{King} \&
  {Pounds}}{2015}]{King2015}
{King} A.,  {Pounds} K.,  2015, \mn@doi [\araa]
  {10.1146/annurev-astro-082214-122316}, \href
  {https://ui.adsabs.harvard.edu/abs/2015ARA&A..53..115K} {53, 115}

\bibitem[\protect\citeauthoryear{{Knowles} et~al.,}{{Knowles}
  et~al.}{2022}]{Knowles2022}
{Knowles} K.,  et~al., 2022, \mn@doi [\aap] {10.1051/0004-6361/202141488},
  \href {https://ui.adsabs.harvard.edu/abs/2022A&A...657A..56K} {657, A56}

\bibitem[\protect\citeauthoryear{{Kormendy} \& {Ho}}{{Kormendy} \&
  {Ho}}{2013}]{KH13.M.Sigma}
{Kormendy} J.,  {Ho} L.~C.,  2013, \mn@doi [\araa]
  {10.1146/annurev-astro-082708-101811}, \href
  {https://ui.adsabs.harvard.edu/abs/2013ARA&A..51..511K} {51, 511}

\bibitem[\protect\citeauthoryear{{Lacy}, {Rawlings}, {Blundell}  \&
  {Ridgway}}{{Lacy} et~al.}{1998}]{Lacy1998}
{Lacy} M.,  {Rawlings} S.,  {Blundell} K.~M.,   {Ridgway} S.~E.,  1998, \mn@doi
  [\mnras] {10.1046/j.1365-8711.1998.01591.x}, \href
  {https://ui.adsabs.harvard.edu/abs/1998MNRAS.298..966L} {298, 966}

\bibitem[\protect\citeauthoryear{{Lintott} et~al.,}{{Lintott}
  et~al.}{2009}]{Lintott09}
{Lintott} C.~J.,  et~al., 2009, \mn@doi [\mnras]
  {10.1111/j.1365-2966.2009.15299.x}, \href
  {https://ui.adsabs.harvard.edu/abs/2009MNRAS.399..129L} {399, 129}

\bibitem[\protect\citeauthoryear{{McNamara} \& {Nulsen}}{{McNamara} \&
  {Nulsen}}{2012}]{McNamara2012}
{McNamara} B.~R.,  {Nulsen} P.~E.~J.,  2012, \mn@doi [New Journal of Physics]
  {10.1088/1367-2630/14/5/055023}, \href
  {https://ui.adsabs.harvard.edu/abs/2012NJPh...14e5023M} {14, 055023}

\bibitem[\protect\citeauthoryear{{Minkowski}}{{Minkowski}}{1958}]{Minkowski1958}
{Minkowski} R.,  1958, \mn@doi [Reviews of Modern Physics]
  {10.1103/RevModPhys.30.1048}, \href
  {https://ui.adsabs.harvard.edu/abs/1958RvMP...30.1048M} {30, 1048}

\bibitem[\protect\citeauthoryear{{Mukherjee}, {Bicknell}, {Sutherland}  \&
  {Wagner}}{{Mukherjee} et~al.}{2016}]{Dipanjan2016}
{Mukherjee} D.,  {Bicknell} G.~V.,  {Sutherland} R.,   {Wagner} A.,  2016,
  \mn@doi [\mnras] {10.1093/mnras/stw1368}, \href
  {https://ui.adsabs.harvard.edu/abs/2016MNRAS.461..967M} {461, 967}

\bibitem[\protect\citeauthoryear{{Nagar}, {Wilson}, {Mulchaey}  \&
  {Gallimore}}{{Nagar} et~al.}{1999}]{Nagar1999}
{Nagar} N.~M.,  {Wilson} A.~S.,  {Mulchaey} J.~S.,   {Gallimore} J.~F.,  1999,
  \mn@doi [\apjs] {10.1086/313183}, \href
  {https://ui.adsabs.harvard.edu/abs/1999ApJS..120..209N} {120, 209}

\bibitem[\protect\citeauthoryear{{Rakshit}, {Stalin}, {Hota}  \&
  {Konar}}{{Rakshit} et~al.}{2018}]{Rakshit2018}
{Rakshit} S.,  {Stalin} C.~S.,  {Hota} A.,   {Konar} C.,  2018, \mn@doi [\apj]
  {10.3847/1538-4357/aaefe8}, \href
  {https://ui.adsabs.harvard.edu/abs/2018ApJ...869..173R} {869, 173}

\bibitem[\protect\citeauthoryear{{Salom{\'e}}, {Salom{\'e}}  \&
  {Combes}}{{Salom{\'e}} et~al.}{2015}]{Salome2015}
{Salom{\'e}} Q.,  {Salom{\'e}} P.,   {Combes} F.,  2015, \mn@doi [\aap]
  {10.1051/0004-6361/201424932}, \href
  {https://ui.adsabs.harvard.edu/abs/2015A&A...574A..34S} {574, A34}

\bibitem[\protect\citeauthoryear{{Shimwell} et~al.,}{{Shimwell}
  et~al.}{2022}]{lotssdr2}
{Shimwell} T.~W.,  et~al., 2022, \mn@doi [\aap] {10.1051/0004-6361/202142484},
  \href {https://ui.adsabs.harvard.edu/abs/2022A&A...659A...1S} {659, A1}

\bibitem[\protect\citeauthoryear{{Swarup}, {Ananthakrishnan}, {Kapahi}, {Rao},
  {Subrahmanya}  \& {Kulkarni}}{{Swarup} et~al.}{1991}]{Swarup1991GMRT2}
{Swarup} G.,  {Ananthakrishnan} S.,  {Kapahi} V.~K.,  {Rao} A.~P.,
  {Subrahmanya} C.~R.,   {Kulkarni} V.~K.,  1991, Current Science, \href
  {https://ui.adsabs.harvard.edu/abs/1991CuSc...60...95S} {60, 95}

\bibitem[\protect\citeauthoryear{{Terrazas} et~al.,}{{Terrazas}
  et~al.}{2020}]{Terrazas2020}
{Terrazas} B.~A.,  et~al., 2020, \mn@doi [\mnras] {10.1093/mnras/staa374},
  \href {https://ui.adsabs.harvard.edu/abs/2020MNRAS.493.1888T} {493, 1888}

\bibitem[\protect\citeauthoryear{{Veilleux}, {Cecil}  \&
  {Bland-Hawthorn}}{{Veilleux} et~al.}{2005}]{Veilleux2005}
{Veilleux} S.,  {Cecil} G.,   {Bland-Hawthorn} J.,  2005, \mn@doi [\araa]
  {10.1146/annurev.astro.43.072103.150610}, \href
  {https://ui.adsabs.harvard.edu/abs/2005ARA&A..43..769V} {43, 769}

\bibitem[\protect\citeauthoryear{{Venturi} et~al.,}{{Venturi}
  et~al.}{2021}]{Venturi2021}
{Venturi} G.,  et~al., 2021, \mn@doi [\aap] {10.1051/0004-6361/202039869},
  \href {https://ui.adsabs.harvard.edu/abs/2021A&A...648A..17V} {648, A17}

\bibitem[\protect\citeauthoryear{{Webster} et~al.,}{{Webster}
  et~al.}{2021}]{Webster2021}
{Webster} B.,  et~al., 2021, \mn@doi [\mnras] {10.1093/mnras/staa3437}, \href
  {https://ui.adsabs.harvard.edu/abs/2021MNRAS.500.4921W} {500, 4921}

\bibitem[\protect\citeauthoryear{{van Breugel} \& {Dey}}{{van Breugel} \&
  {Dey}}{1993}]{vanBreugel1993}
{van Breugel} W. J.~M.,  {Dey} A.,  1993, \mn@doi [\apj] {10.1086/173103},
  \href {https://ui.adsabs.harvard.edu/abs/1993ApJ...414..563V} {414, 563}

\makeatother
\end{thebibliography}





\bsp	
\label{lastpage}
\end{document}